%% file: prop_analysis.tex
\documentclass[copyright,creativecommons]{eptcs}
\providecommand{\event}{WWV 2015} % Name of the event you are submitting to

\usepackage{breakurl}             % Not needed if you use pdflatex only.
\usepackage{amsmath}
\usepackage{amsfonts}
\usepackage{amssymb}
\usepackage{multicol}
\usepackage[textsize=footnotesize,colorinlistoftodos,textwidth=1.9cm]{todonotes}
\usepackage{url}
\usepackage{soul}
\usepackage{acronym}
\usepackage{hyperref}
\usepackage{color}
\usepackage{proof}
\usepackage{wasysym}

\newcommand{\facpl}{FACPL}

\title{On Properties of Policy-Based Specifications\thanks{This work has been partially sponsored by the Italian MIUR PRIN project CINA (2010LHT4KM).}}
\author{Andrea Margheri
\institute{Universit\`a degli Studi di Firenze}
\email{andrea.margheri@unifi.it}
\institute{Universit\`a di Pisa}
\email{margheri@di.unipi.it}
\and
Rosario Pugliese
\institute{Universit\`a degli Studi di Firenze}
\email{rosario.pugliese@unifi.it}
\and
Francesco Tiezzi
\institute{Universit\`a di Camerino}
\email{francesco.tiezzi@unicam.it}
}

\def\titlerunning{On Properties of Policy-Based Specifications}
\def\authorrunning{A. Margheri, R. Pugliese \& F. Tiezzi}

\begin{document}
\maketitle

\input{tex/macro}

\begin{abstract}
The advent of large-scale, complex computing systems has dramatically increased the difficulties of securing accesses to systems' resources. To ensure confidentiality and integrity, the exploitation of access control mechanisms has thus become a crucial issue in the design of modern computing systems. Among the different access control approaches proposed in the last decades, the policy-based one permits to capture, by resorting to the concept of attribute, all systems' security-relevant information and to be, at the same time, sufficiently flexible and expressive to represent the other approaches. In this paper, we move a step further to understand the effectiveness of policy-based specifications by studying how they permit to enforce traditional security properties. To support system designers in developing and maintaining policy-based specifications, we formalise also some relevant properties regarding the structure of policies. By means of 
a case study from the banking domain, we present real instances of such properties and outline an approach towards their automatised verification.
\end{abstract}

\input{tex/intro}
\input{tex/facpl}
\input{tex/security_properties}
\input{tex/struct_prop}
\input{tex/constraint}

\input{tex/relatedworks}
\input{tex/conclusion}

\bibliographystyle{eptcs}
\bibliography{biblio}

\end{document}

%% file: tex/macro.tex
% !TEX root =  ../prop_analysis.tex

\definecolor{mygrey}{gray}{0.75}
\definecolor{grigiochiaro}{gray}{0.9}
\definecolor{grigiomoltochiaro}{gray}{0.97}
\definecolor{verde}{rgb}{0,250,0}

\newcommand{\typeOne}{\texttt{TYPE\_1}}
\newcommand{\typeTwo}{\texttt{TYPE\_2}}

\newcommand{\ALGOXML}[3]
{
  \lstset{
   %    basicstyle=\tiny\sffamily,
    basicstyle=\scriptsize\sffamily,
    stringstyle=\color{red},
    commentstyle=\color{green},
    keywordstyle=\color{blue}\bfseries\underbar,
    caption={#1},
    label={#2},
    frame={tb},
    numbers=left,
    numberstyle=\tiny,
    numbersep=5pt,
    tabsize=4,
    language=XML,
    showtabs=false,
    showstringspaces=false,
    identifierstyle=,
    breaklines,
    literate={~=}{{$\neq$}}2{<=}{{$\leq$}}2{>=}{{$\geq$}}2{&}{{$\&$}}2,
    %    backgroundcolor=\color{grigiochiaro}
    backgroundcolor=\color{grigiomoltochiaro}
    }
  \lstinputlisting{src_xml/#3}
}

\newcommand{\ALGOInline}[3]
{
  \lstset{
   %    basicstyle=\tiny\sffamily,
    basicstyle=\scriptsize\ttfamily,
    stringstyle=\color{red},
    commentstyle=\color{verde},
    keywordstyle=\color{blue}\bfseries\underbar,
    caption={#1},
    %label={#2},
    frame={tb},
    numbers=none,
    numberstyle=\tiny,
    numbersep=5pt,
    tabsize=4,
    language=Java,
    showtabs=false,
    showstringspaces=false,
    identifierstyle=,
    breaklines,
    literate={~=}{{$\neq$}}2{<=}{{$\leq$}}2{>=}{{$\geq$}}2{&}{{$\&$}}2,
    %    backgroundcolor=\color{grigiochiaro}
    backgroundcolor=\color{grigiomoltochiaro}
    }
  \lstinputlisting{src/#3}
}

\newcommand{\ALGOFPL}[3]
{
  \lstset{
   %    basicstyle=\tiny\sffamily,
    basicstyle=\scriptsize\ttfamily,
    stringstyle=\color{red},
    commentstyle=\color{verde},
    keywordstyle=\color{blue}\bfseries\underbar,
    caption={#1},
    label={#2},
    frame={tb},
    numbers=none,
    numberstyle=\tiny,
    numbersep=5pt,
    tabsize=4,
    language=Java,
    showtabs=false,
    showstringspaces=false,
    identifierstyle=,
    breaklines,
    literate={~=}{{$\neq$}}2{<=}{{$\leq$}}2{>=}{{$\geq$}}2{&}{{$\&$}}2,
    %    backgroundcolor=\color{grigiochiaro}
    backgroundcolor=\color{grigiomoltochiaro}
    }
  \lstinputlisting{src/#3}
}

\newcommand{\ie}{i.e.~}
\newcommand{\eg}{e.g.~}

\newcommand{\modif}[1]{{\color{red}#1}}

%---------------------------------------
%ACRONYM
%---------------------------------------
%\newcommand{\xacml}{\ac{XACML}}
%\newcommand{\facpl}{\ac{FACPL}}

\newcommand{\pdp}{\ac{PDP}}
\newcommand{\pep}{\ac{PEP}}
\newcommand{\pap}{\ac{PAP}}
\newcommand{\pip}{\ac{PIP}}

\newcommand{\dsl}{\ac{DSL}}
\newcommand{\ide}{\ac{IDE}}

%---------------------------------------

\newcommand{\ccvar}[1]{\textsf{#1}} %constrained variables

\newcommand{\arr}[1]{\langle #1 \rangle} % notation for array/tuple
\newcommand{\set}[1]{\{#1\}} % brackets for sets

\newcommand{\und}{\_} % underscore symbol used to denote an unspecified argument

\newcommand{\myfont}[1]{\mathrm {#1}} % normal font for math environment
\newcommand{\true}{\x{true}} % the boolean true
\newcommand{\false}{\x{false}} %the boolean false

\newcommand{\substl}{\sigma} % notation for substitution
\newcommand{\substi}[1]{\{#1\}} % bracket for susbtitution
\newcommand{\assoc}[2]{#1 \mapsto #2} % assocition variable-value
\newcommand{\length}[1]{\mid\! #1 \!\mid} % lungth of a substitution

\newcommand{\per}{\times} % operazione prodotto di un c-semiring
\newcommand{\zero}{0} % elemento 0 del c-semiring
\newcommand{\uno}{1} % elemento 1 del c-semiring

\newcommand{\isCons}[1]{\mathit{isCons}(#1)} % operazione prodotto di un c-semiring
\newcommand{\entail}[2]{#1 \vdash #2}
\newcommand{\ccunion}{\uplus}
\newcommand{\ccmarked}[1]{#1^{\vdash}} % vincolo marchiato

\newcommand{\Sep}{\ \, \mid\ \,}

\newcommand{\x}[1]{{\sf #1}}
\newcommand{\pdpPol}[2]{\{ #1 \,  \, #2 \}}
\newcommand{\xacmlPol}[2]{\{ #1 \, ; \, #2 \}}
\newcommand{\pepPol}[1]{\x{PepAlg:} \{ #1 \}}

%%%%
\newcommand{\eval}[1]{\langle \! \langle #1 \rangle \! \rangle}
\newcommand{\evalS}[1]{[ \! [ #1 ] \! ]}

%%%%%

%-----------------------------------
%algorithm
%-----------------------------------
\newcommand{\onlyOneApp}{\x{only}\textrm{-}\x{one}\textrm{-}\x{applicable}}
\newcommand{\firstApp}{\x{first}\textrm{-}\x{applicable}}
\newcommand{\denyOver}{\x{deny}\textrm{-}\x{overrides}}
\newcommand{\permitOver}{\x{permit}\textrm{-}\x{overrides}}
\newcommand{\ordDenyOver}{\x{ordered}\textrm{-}\x{deny}\textrm{-}\x{overrides}}
\newcommand{\ordPermitOver}{\x{ordered}\textrm{-}\x{permit}\textrm{-}\x{overrides}}
\newcommand{\based}{\x{base}}
\newcommand{\debugAlg}{\x{debug}}
\newcommand{\denyBiased}{\x{deny}\textrm{-}\x{biased}}
\newcommand{\permitBiased}{\x{permit}\textrm{-}\x{biased}}
\newcommand{\permitUnless}{\x{permit}\textrm{-}\x{unless}\textrm{-}\x{deny}}
\newcommand{\denyUnless}{\x{deny}\textrm{-}\x{unless}\textrm{-}\x{permit}}
\newcommand{\weakCon}{\x{weak}\textrm{-}\x{consensus}}
\newcommand{\strongCon}{\x{strong}\textrm{-}\x{consensus}}

\newcommand{\all}{\x{all}}
\newcommand{\greedy}{\x{greedy}}

%-------------------------------------

%-------------------------------------
%policy/policySet/rule syntax
%-------------------------------------

\newcommand{\polSet}[4]{{\bf{\{}}#1{\,}\x{target:}\,#2 {?}\, \x{policies:}#3^{+} {\,}\x{obl:}#4^{*}\, {\bf{\}}}}

\newcommand{\polSetE}[4]{{\bf{\{}}#1 \, \x{target:}\,#2 \, \x{policies:}#3 \, \x{obl:}#4 \, {\bf{\}}}}

\newcommand{\pol}[4]{{\bf{\langle}}#1{\,}\x{target:}\,#2 {?}{\,}\, \x{rules:}#3^{+} {\,}\x{obl:}#4^{*}{\,}{\bf{\rangle}}}

\newcommand{\polE}[4]{{\bf{\langle}}#1{\,}\x{target:}\,#2 {\,}\, \x{rules:}#3 {\,}\x{obl:}#4 {\,}{\bf{\rangle}}}

\newcommand{\obl}[1]{\, PepAction( #1^* ) }

%---------------------------------------
% Separator
%-------------------------
%\newcommand{\policy}{Policy}
%\newcommand{\Rule}{Rule}
%\newcommand{\Target}{Target}
%\newcommand{\MatchId}{MathId}
%\newcommand{\Obligation}{Obligation}
%\newcommand{\Value}{Value}
%\newcommand{\Name}{Name}
%
%\newcommand{\policy}{\pi}
%\newcommand{\Rule}{r}
%%\newcommand{\Target}{\tau}
%\newcommand{\Target}{\tau}
%\newcommand{\MatchId}{f}
%\newcommand{\Obligation}{o}
%\newcommand{\Value}{\mathit{pv}}
%\newcommand{\Name}{\mathit{sn}}
%
%\newcommand{\Pa}{\alpha}
%\newcommand{\Ra}{\alpha}

\newcommand{\allow}{\vdash}
\newcommand{\valt}[2]{\mbox{$[\![ \: #1 \: ]\!]_{#2}$}}

\newcommand{\policySetBegin}{{\bf{\{}}}
\newcommand{\policyBegin}{{\bf{\langle}}}
\newcommand{\policySep}{\,\bf{\ }\,}
\newcommand{\policiesBegin}{\x{policies:}\,}
\newcommand{\targetBegin}{\x{target:}\,}
\newcommand{\targetEnd}{\,{\bf{\ }}}
\newcommand{\policiesEnd}{\,{\bf{\ }}}
\newcommand{\rulesBegin}{\x{rules:}\,}
\newcommand{\rulesEnd}{\,{\bf{\ }}}
\newcommand{\policyEnd}{{\bf{\rangle}}}
\newcommand{\policySetEnd}{{\bf{\}}}}
\newcommand{\conditionBegin}{\x{condition:}\,}
\newcommand{\conditionEnd}{\,{\bf{\ }}}
\newcommand{\oblsBegin}{\x{obl:}\,}
\newcommand{\oblsEnd}{\bf{\ }\,}

\newcommand{\oblBegin}{{\bf{[}}}
\newcommand{\oblEnd}{{\bf{]}}}

\newcommand{\match}[3]{#1{\bf(}#2{\bf,}#3{\bf)}}
\newcommand{\ruleOpt}[1]{{\bf{(}}#1{\bf{)}}}
\newcommand{\oblOpt}[1]{{\bf{(}}#1{\bf{)}}}
%------------------------------
%Decisions
%------------------------------
\newcommand{\permit}{\x{permit}}
\newcommand{\deny}{\x{deny}}
\newcommand{\notApp}{\x{not}\textrm{-}\x{applicable}}
\newcommand{\indet}{\x{indeterminate}}

\newcommand{\oblig}{\x{obligations}}
\newcommand{\Match}{\x{match}}
\newcommand{\noMatch}{\x{no}\textrm{-}\x{match}}

\newcommand{\streq}{\x{equal}}

%-----------------------
%contextReq
%-----------------------
\newcommand{\contextReq}[1]{\x{request:} \, \{ \textit{#1} \} }
\newcommand{\contextReqBegin}{\x{request:} \, \{ }
\newcommand{\contextReqEnd}{ \} }
\newcommand{\contextResp}[1]{\x{response:} \, \{ \textit{#1} \} }
\newcommand{\contextRespPDPBegin}[1]{\x{AD:} \,  }
\newcommand{\contextRespPEPBegin}[1]{\x{ED:} \,  }
\newcommand{\contextRespEnd}{ \} }
\newcommand{\contextSep}{\, ; \,}
\newcommand{\context}[2]{\contextReq{#1}\contextSep\contextResp{#2}}
\newcommand{\contextOpt}[2]{\contextReq{#1}[\contextSep\contextResp{#2}\,]}
\newcommand{\attribute}[2]{( #1 , #2 )}
\newcommand{\result}[1]{( #1 )}

\newcommand{\reqsetExt}[2]{\{#1 \,\mid\, #2\}}

\newcommand{\mEl}{m}
\newcommand{\mEval}[2]{#2 \models #1}

%-------------------------------
%Commands for semantics
%-------------------------------
\newcommand{\req}{r}
\newcommand{\Req}{R}
\newcommand{\reqsetAll}{R_{all}}

%%%%%%%%%%%%%%%%%%%%%%%%%%%%%%%%%%%%%%%%%%%%%%%%%%%%%%%%%%%%%%
\newcommand{\excpt}{\perp}
\newcommand{\err}{\x{error}}

%%%%%%%%%%%%%%%%%%%%%%%
%FORMAL SEMANTICS
%%%%%%%%%%%%%%%%%%%%%%%

%Auxiliary Semantic Functions and notations
\newcommand{\pepSemR}[1]{(\!( #1 )\!)}
\newcommand{\concat}{{\bullet}}
\newcommand{\subobeff}[2]{#1\!\!\mid_{#2}} %sub sequence of obligations

%Name of Semantic Function
\newcommand{\denSemF}[1]{\mathcal{ #1 }}

%Semantic Brakets
%\newcommand{\denSem}[2]{[\![ #1 ]\!]_{#2}}
\newcommand{\denSem}[2]{[\![ #1 ]\!] #2}

%Named Semantics Functions
\newcommand{\policySemF}[2]{\denSemF{P}\denSem{#1}{#2}}
\newcommand{\algSem}[2]{\denSemF{A}\denSem{#1}{#2}}
\newcommand{\exprSem}[2]{\denSemF{E}\denSem{#1}{#2}}
\newcommand{\oblSem}[2]{\denSemF{O}\denSem{#1}{#2}}
\newcommand{\oblSemS}[2]{\denSemF{OS}\denSem{#1}{#2}}
\newcommand{\pdpSem}[2]{\denSemF{P}dp\denSem{#1}{#2}}
\newcommand{\pepSem}[2]{\denSemF{E}A\denSem{#1}{#2}}
\newcommand{\pasSem}[1]{\denSemF{P}as\denSem{#1}{}}
\newcommand{\reqSemS}[2]{\denSemF{R}\denSem{#1}{#2}}  %two arguments, \name as second argument
\newcommand{\reqSem}[1]{\denSemF{R}\denSem{#1}{}} 

\newcommand{\define}{\triangleq}

%Generic Element of each Syntactic Category
\newcommand{\expr}{\mathit{expr}}
\newcommand{\effect}{\mathit{e}}
\newcommand{\ob}{\mathit{o}}
\newcommand{\obType}{\mathit{t}}
\newcommand{\fo}{\mathit{f\!o}}
\newcommand{\foS}{\mathit{f\!o}^*}
\newcommand{\rSyntax}{\mathit{req}} %element of syntactic category Request
\newcommand{\policy}{\mathit{p}}
\newcommand{\algSyntax}{\mathit{a}}
\newcommand{\pdpRes}{\mathit{res}}
\newcommand{\dec}{\mathit{dec}} %decision
\newcommand{\double}{\mathit{d}} %double
\newcommand{\extVal}{\mathit{w}} % Obligations and table of expression semantics
\newcommand{\enfAlg}{\mathit{ea}}
\newcommand{\pdpSyntax}{\mathit{pdp}}
\newcommand{\name}{\mathit{n}}
\newcommand{\val}{\mathit{v}}
\newcommand{\pepAction}{\mathit{pepAct}}

\newcommand{\algNT}{\mathit{Alg}} %used in syntax section
\newcommand{\algName}{\mathsf{alg}} % used in combining algorithms semantic section

%Semantics of Combiing Algorithms
\newcommand{\alg}[1]{\mathsf{alg}_{#1}}
\newcommand{\algD}{\alg{\delta}}
\newcommand{\algOp}{\otimes \mathsf{alg}}
\newcommand{\algOpAlg}[1]{\otimes #1}

\newcommand*\lfrac[2]{{}_{#1}\!\backslash\!^{#2}} % table combining algorithm

%%%%%%%%%%%%%%%%%%%%%%%%%%%%%%%%%%%%%%%%%%%%%%%%%%%%%%%%%%%%%%%%

%%%%%%%%%SEMANTICA UTILIZZATA NEL PAPER WWV%%%%%%%%%%%
\newcommand{\policySem}[1]{[\![ #1 ]\!]}
%%%%%%%%%%%%%%%%%%%%%%%%%%%%%%%%%%%%%%%%%%%

\newcommand{\roleOne}{\emph{assistant}}
\newcommand{\roleTwo}{\emph{officier}}

\newcommand{\doc}{loanDoc}

\newcommand{\submit}{\emph{submit}}
\newcommand{\approve}{\emph{approve}}

\newcommand{\pobreq}[1]{\overline{#1}}
\newcommand{\mobreq}[1]{\underline{#1}}

%% file: tex/intro.tex
% !TEX root =  ../prop_analysis.tex

\section{Introduction}
\label{sec:intro}

The ever increasing diffusion of the Internet and the Web has fostered the development of large-scale, complex computing systems. These modern distributed systems, that are pervading our everyday life, produce and exploit a huge amount of data that are readily available through the underlying network platforms. Given their importance and societal impact, it is of paramount importance to ensure that data is accessed in a controlled way and that these systems behave in a secure way, e.g.~not to compromise sensitive data. For achieving this objective, some major challenges come from the fact that the operating environment is highly dynamic and open, the involved entities are heterogeneous and possibly untrusted, the interactions are complex and unpredictable, and the control is distributed. 

In this setting, we believe that policy-based specifications can be used to regulate the behaviour of entities relatively to the access to shared resources, thus ensuring systems \emph{security}. \emph{Policies}, that is sets of declarative rules expressing what can(not) be done in a system, are indeed high-level abstractions that can be used to define various aspects of systemsÕ behaviour. In particular, a \emph{security policy} is a statement that defines in which states a system is considered secure. A system is \emph{secure} if starting from a secure state it cannot enter a nonsecure one while computation progresses. The security of a state depends on the behaviours the system exposes and, hence, on which guarantees a security policy managing and controlling the system ensures. The enforcement of such a policy relies on a combination of various approaches, ranging, e.g., from cryptography to access control, according to features and specificity of the controlled system.

We focus on \emph{access control}, usually considered the first line of defence in protection of computer systems, networks, and information. Access control is a broad field that covers several different approaches that enable the protection of systems by restricting physical and logical access rights of (authenticated) subjects to shared resources. In practice, these approaches establish if a subject's request to access a resource should be permitted or denied according to some given access control rules. 

Since their original introduction in the context of operating systems, to the more recently conceived ones for modern distributed applications, many approaches for access control have been proposed in the literature. Traditional approaches are based on the identity of the subject, either directly \,--\,e.g., Access Control Matrix~\cite{Lampson74,GD72} and its variants Capability Lists and Access Control Lists\,--\, or through predefined attributes, such as roles or groups assigned to that subject --\,e.g., Role-Based Access Control (RBAC~\cite{rbac}). In our frame of reference, these approaches are cumbersome to manage and not sufficiently expressive, given the need to associate access rights to the requester qualifiers of identity, groups, and roles that can change frequently and could not be known in advance. To overcome scalability problems of these traditional access control approaches, an alternative is to use Attribute-Based Access Control (ABAC~\cite{nistsurvey}). Here, the authorisation decision is based on \emph{attributes}, which represent arbitrary information exposed by the system, subject, action, object, or the authorisation context itself that is relevant to the rules at hand. Thus, ABAC permits defining fine-grained, flexible and context-aware access control policies, and fosters systems integration, as attributes can be retrieved from different information systems.  Attribute-based access control rules are typically hierarchically structured and paired with strategies for automatic treatment of conflicting decisions and errors. These structured specifications are called \emph{policies}; from this name derives the terminology Policy-Based Access Control (PBAC), sometimes used in the literature in place of ABAC.

Approaches to access control can be classified also with respect to other features. For example, if we consider resource ownership then we can distinguish between \emph{discretionary} access control (DAC), where subjects may decide who can access their own resources, i.e. the access control is at the discretion of the owner, and \emph{mandatory} access control (MAC),  where the system decides who is allowed to access any resource. In this respect, the access control matrix better fits the DAC approach, while RBAC and ABAC can be used both for the MAC and DAC approaches. 
To conclude this overview of the relevant access control approaches, on the base of the results in~\cite{JinKS12}, we can say that ABAC is sufficiently expressive to represent in an uniform way all the other approaches. 
 
Controlling accesses to system resources concerns the three main security principles of \emph{confidentiality}, \emph{integrity}, and \emph{availability}. Specifically, confidentiality refers to the assurance on non-disclosure of sensitive resources to unauthorised subjects; integrity to the protection of resources from being altered by unauthorised subjects; and availability to the enablement of the effective use of resources by authorised subjects. 
As instantiations of these general principles, many security properties have been introduced and studied (e.g., the Bell-LaPadula~\cite{Bell76} and Biba~\cite{Biba} models). However, enforcing such properties by means of access control policies is a tricky task. In fact, the hierarchical structure of policies, the presence of conflict resolution strategies and the intricacies deriving from the many involved controls do not permit to easily check whether a given security property is properly enforced. Therefore, in our work we consider a general instance of the ABAC approach, i.e. the \facpl\ language~\cite{FACPLTR}, and study in details a set of relevant security properties, presenting how they can be rendered in terms of policy-based specifications.

Policy-based specifications are formed by multiple rules and policies, and to characterise the relationships with the behaviours they enforce, various properties on the structure of policies have been proposed in the literature (e.g., change-impact analysis~\cite{FislerKMT05} and redundancy minimisation~\cite{GuarnieriNMM13}). The approaches used for defining and verifying these properties are different and cannot be uniformly represented. Therefore, in our work we focus on a set of relevant structural properties and propose a uniform formalisation in terms of the \facpl\ semantics. 

Furthermore, for providing a concrete support to the verification of both security and structural properties, we outline a constraint-based approach enabling automated verification by means of constraint solver software tools. At the time of writing, this constraint-based analysis of policies is under development; thus, in this paper, we just present main features and strengths of the approach.

\smallskip
\noindent 
\emph{Case Study}. We consider a scenario from the banking domain where access control policies are used for managing money loan activities. We assume that a costumer willing to borrow money from a bank has to fill out a \emph{loan request document}, possibly aided by a bank clerk. Once the document is finalised, the clerk submits it for approval; if the document is approved, the loan is granted. Each loan request is associated with a confidentiality level, so that a high confidential loan request has to be managed only by highly-trusted clerks. Notably, the \emph{approve} (resp., \emph{submit}) actions of loan request documents are carried out by clerks that, for safety reasons, cannot do \emph{submit} (resp., \emph{approve}) actions. 

The case study is addressed throughout the paper. First, we introduce the structure of policies and access requests,  then, by characterising a set of relevant security properties, we define some policies that can be used for managing different aspects of the money loan activities.

\smallskip
\noindent 
\emph{Outline of the rest of the paper}. Section~\ref{sec:facpl} briefly reports main features of policy-based languages and introduces the \facpl\ policy language. Section~\ref{sec:sec_prop} presents the representation in terms of policy-based specifications and the formalisation of a set of security properties, while Section~\ref{sec:struct} addresses policies' structural properties. The verification approach, together with our proposal towards an automated tool support, is sketched in Section~\ref{sec:tool}. Finally, Section~\ref{sec:relwork} reviews more strictly related work and Section~\ref{sec:concl} concludes by touching upon directions for future work. Background definitions and concepts on computer security used in rest of the paper are based on the well-known text books~\cite{Bishop, Gollmann0025849}.

%% file: tex/facpl.tex
% !TEX root =  ../prop_analysis.tex

\section{A Policy Language}
\label{sec:facpl}

Policy languages for access control provide high-level abstractions for the specifications of declarative sets of access control rules. Specifically, these languages allow systems' designers to express structured sets of attribute-based positive (resp. negative) rules granting (resp. forbidding) the access to systems' resources. In this section, by informally introducing (a light version of) the policy language \facpl~\cite{FACPLTR}, we detail all the typical features of access control specifications. The authorisation process that is pursued to authorise or forbid an access request is outlined by means of a simple example.

\subsection{Syntax and Informal Semantics of \facpl}
\label{sec:policySyntax}

The syntax of a light version of \facpl\ is reported in Table~\ref{tab:facpl_syntax}. It is given trough EBNF-like grammars, where as usual the symbol $?$ indicates optional items and $+$ indicates non-empty sequences of items.

The top-level term is a $PDP$, which is defined by a sequence of policies $\mathit{Policy}^{+}$ and an algorithm $Alg$ for combining the results of the evaluation of these policies.

\begin{table}[!t]
\caption{Syntax of a light version of \facpl}
\label{tab:facpl_syntax}
\centering
\small
$
\begin{array}{@{\,}r@{\ \ }r@{\ }r@{\ \ }l@{\ }}
\hline
&&&\\[-.2cm]
{\textbf{Policy Decision Points}} &
\mathit{PDP} & ::= & \pdpPol{\algNT\ }{\x{policies:} \, \mathit{Policy}^{+}}
\\[.2cm]
{\textbf{Combining algorithms}} &
\algNT & ::= & \permitOver \Sep \denyOver \Sep \denyUnless \\
&& \mid &
\permitUnless \Sep \firstApp \Sep \onlyOneApp \\
&&\mid &
\weakCon \Sep \strongCon 
\\[.2cm]
{\textbf{Policies}} &
\mathit{Policy} & ::= &
\ruleOpt{\mathit{Effect}\ \ \mathtt{target:} \, Expr  
} \\
&& \mid &
\{ \algNT\ \ \mathtt{target:} \, Expr\ \ \mathtt{policies:} \, \mathit{Policy}^{+}  
\}
\\[.2cm]
{\textbf{Effects}} &
\mathit{Effect} & ::= & \permit \Sep \deny
\\[.2cm]
\textbf{Expressions}&
\mathit{Expr} & ::= &
\mathit{Name} \Sep \mathit{Value}  \Sep
\x{and(\mathit{Expr}, \mathit{Expr})} \Sep \x{or(\mathit{Expr}, \mathit{Expr})} \Sep \x{not(\mathit{Expr})} \\
& & \mid &
 \x{equal(\mathit{Expr},\mathit{Expr})}  \Sep \x{in}(\mathit{Expr}, \mathit{Expr}) \Sep  \x{greater}\textrm{-}\x{than(\mathit{Expr},\mathit{Expr})} \\
& & \mid & \x{add(\mathit{Expr} ,\mathit{Expr} )} \Sep \x{subtract(\mathit{Expr} ,\mathit{Expr} )} \\
& & \mid & \x{divide(\mathit{Expr} ,\mathit{Expr} )} \Sep \x{multiply(\mathit{Expr} ,\mathit{Expr} )} \\[.2cm]
\textbf{Attribute Names} & 
\mathit{Name} & ::= & \mathit{Identifier}/\mathit{Identifier} \\[.2cm]
\textbf{Literal Values} &
\mathit{Value} & ::= & \x{true} \Sep \x{false} \Sep \mathit{Double} \Sep \mathit{String} \Sep \mathit{Date}
\\[.4cm]
{\textbf{Access Requests}} &
\mathit{Request} & ::= & {\attribute{\mathit{Name}}{\mathit{Value}}}^{+}
\\[.1cm]
\hline
\end{array}
$
\end{table}

A \emph{policy} can be a basic authorization \emph{rule} $\ruleOpt{\mathit{Effect}\ \ \mathtt{target:} \, Expr}$ or a \emph{policy set} $\{ \algNT\ \ \mathtt{target:} \, Expr\ \ \mathtt{policies:} \, \mathit{Policy}^{+}  \}$ collecting rules and (other) policy sets, so that it is possible to define hierarchical policies. A rule specifies an $\mathit{Effect}$, i.e. $\permit$ or $\deny$, indicating the rule-writer's intended consequence of a successful evaluation for the rule, and a \emph{target}, i.e.~an expression $Expr$ defining the applicability of the rule to a request. A \emph{policy} instead specifies a target, a sequence of contained elements, i.e. rules or policies themselves, and an algorithm $\algNT$ for combining the results of the evaluation of these contained elements.

A \emph{combining algorithm} implements a strategy for resolving conflicts among the decisions resulting from the evaluation of a collection of rules/policies, e.g. whenever both decisions $\permit$ and $\deny$ are returned. We report below the strategies implemented by some of these algorithms.
\begin{itemize}
\item $\permitOver$: if the processing of an element returns $\permit$, then the result is $\permit$, i.e., $\permit$ takes precedence over any other decision. Instead, if at least one element returns $\deny$ and all others return $\notApp$ or $\deny$, then the result is $\deny$. If all elements return $\notApp$, then the result is $\notApp$. In the remaining cases, the result is $\indet$.
\item $\denyUnless$: similarly to $\permitOver$, $\permit$ takes precedence over $\deny$, but it never returns $\notApp$ or $\indet$, which are instead evaluated as $\deny$.
\item $\strongCon$: it returns $\permit$ (resp., $\deny$) only if all elements return $\permit$ (resp., $\deny$). If all elements return $\notApp$ then the result is $\notApp$. Otherwise, it returns $\indet$.
\end{itemize}

A \emph{target} is an expression indicating the access requests to which a policy applies. 
\emph{Expressions} are built from attribute names and \emph{literal} values, \ie booleans, doubles, strings, and dates, by using standard operators. As usual, string values are written as sequences of characters delimited by double quotes. For simplicity sake, the expressions syntax does not take types explicitly into account; however, the semantics of expressions returns an error if the arguments of operations have incorrect types. The latter can be anyway managed by policies by resorting to appropriate combining algorithms. 

An $\mathit{attribute\ name}$ indicates the value of an attribute within an access request to authorise. Attributes are expressed in terms of pairs name-value, where names are structured in the form $\x{cat}/\x{att}$, with $\x{cat}$ standing for a category name (as, e.g., \x{subject}, \x{resource}, \x{action}) and $\x{att}$ for a specific name (as, e.g., \x{id} and \x{role}). For example, the structured name $\x{subject/role}$ represents the value of the attribute $\x{role}$ within the category $\x{subject}$. 

An \emph{access request} represents a subject willing to execute an action that has to be authorised. This request holds all the attributes relevant for taking the authorisation decision, such as the information of the subject originating the request and that of the requested action. 

For instance, if we consider the banking case study introduced in Section~\ref{sec:intro}, a request by a subject named $clerk1$ with the assigned role $\roleOne$, i.e. the clerk assisting a customer, that wants to $read$ the resource $\doc$, i.e. the loan request document, is as follows:
$$
(\x{subject/id}, {clerk1}) \ \ (\x{subject/role}, {\roleOne})\ \  (\x{resource/id}, {\doc}) \ \ (\x{action/id}, read) 
$$
All the other case study actions are dealt with similarly. For example, the $\submit$ action of a $\doc$ by an $\roleOne$ clerk, or the $\approve$ action of a $\doc$ by a clerk with the assigned role $\roleTwo$, are simply represented by means of a different instantiation of the previous set of attribute names.

The evaluation of a request with respect to a policy results in one decision among $\permit$, $\deny$, $\notApp$, and $\indet$. The meaning of the first two decisions is obvious (i.e., granting and forbidding the access, respectively), while the third means that there is no policy that applies to the request and the fourth means that some errors occur in the evaluation.

By way of example, to allow read accesses to the resource $loanDoc$ only to subjects with the assigned role $\roleOne$, we might use the following policy
$$
\begin{array}{l}
\{ \denyUnless \\
\ \ \mathtt{target}: \x{equal}(\x{resource/id},
{\doc})\\
\ \ \mathtt{policies}:
 (\permit\ \ \mathtt{target}:  \x{equal}(\x{action/id}, read)\ \x{and}\ \x{equal}(\x{subject/role}, {\roleOne}))
\}
\end{array}
$$
The evaluation of the previous request with respect to the policy above starts by evaluating the policy's target, i.e. the boolean expression after the first keyword $\mathtt{target}$. Since the request satisfies the \x{equal} comparison function, the evaluation carries on with the enclosed rule. The rule's target is satisfied as well and the decision $\permit$ is returned. Then, the combining algorithm applies to the resulting set of decisions, which in this case only contains the $\permit$ one, and returns the final decision for the policy, i.e.~$\permit$. Notice that the policy does not authorise requests not exposing the value {$\roleOne$} as a \x{role} and that the remaining requests are granted only if they ask for $read$ operations. Also notice that if the policy's target does not apply to a request, then $\notApp$ is immediately returned without evaluating the enclosed rule and applying the combining algorithm.

\subsection{A glimpse of the FACPL Formal Semantics}

In this section, we briefly outline the formal semantics of \facpl\ (we refer the interested reader to~\cite{FACPLTR} for a full account). The semantics is defined by following a denotational approach, which means that
\begin{itemize}
\item we introduce some semantic functions mapping each \facpl\ syntactic construct to an appropriate \emph{denotation}, that is an element of a semantic domain representing the meaning of the construct;
\item the semantic functions are defined in a \emph{compositional} way, so that the semantic of each construct is formulated as a function of the semantics of its immediate sub-constructs.
\end{itemize}
To this purpose, for each \facpl\ syntactic category, we specify the semantic domain into which the syntactic constructs map and define the semantic function $\policySem{\ }$ by giving its domain and codomain, and by using semantic clauses to specify, inductively on the syntactic constructs, how the function acts on each construct. Thus, if $P$ stands for a \facpl\ policy, $\policySem{P}{\req}$ corresponds to the decision resulting from the application of the semantic function to (the syntactic object) $P$ and (the semantic object) $\req$ representing an access request.

A \facpl\ request, in order to be evaluated, is represented in its functional form. This is a function $\req$ belonging to the set $R \define \mathit{Name} \rightarrow (\mathit{Value} \cup 2^{\mathit{Value}}\cup \{ \excpt \})$ containing all those total functions mapping attribute names (i.e., the structured names in the syntactic domain $\mathit{Name}$) to either values, or set of values, or the special value $\excpt$ (modelling the fact that an attribute name is missing).

The semantics of a policy is then a function that, given a request, returns an authorisation decision. Formally, it is a function of the form $R \rightarrow \mathit{Decision}$, where $Decision$ corresponds to the semantic (and syntactic) domain of authorisation decisions. To define the semantics of policies we use two clauses, one deals with rules, the other one with policies. For a generic rule $\ruleOpt{\effect\ \ \mathtt{target:} \, \expr\,}$, its semantics is given by the following clause:
$$
\policySem{ \ruleOpt{\effect\ \ \mathtt{target:} \, \expr\,} }{\req} \, = \!\!\!\!
\begin{array}{l}
    \left\{
    \begin{array}{l@{\ \ }l}
    \effect &  \mathtt{if} \ \policySem{\expr}\req=\true\ 
    \\[.1cm]
    \notApp & \mathtt{if} \ \policySem{\expr}\req=\false \ \vee\ \policySem{\expr}\req=\ \excpt\\[.1cm]
    \indet &  \mathtt{otherwise}\\
    \end{array}
    \right.
\end{array}
$$
where $\policySem{\expr}\req$ is the value returned by evaluating the target expression $\expr$ with respect to the request $\req$. Thus, the rule's decision is returned when the target evaluates to $\true$, which means that the rule applies to the request. Otherwise, it could be the case that the rule does not apply to the request, \ie when the target evaluates to $\false$ or to $\excpt$ (which means that the target is an attribute name missing in the request), or that an error has occurred while evaluating the target.

Since the clause for policies relies on the semantics of combining algorithms, we first introduce it. For each combining algorithm, we use a two-dimensional matrix that, given two decisions, calculates the resulting combined one; then, by means of an iterative application of this matrix, we can define the decision returned by the algorithm when given as input a sequence of decisions (each resulting from the evaluation of a policy or a rule). For example, Table~\ref{tab:combiningStrategies_pOver2} reports the matrix for $\permitOver$. Notably, when a matrix takes into account the order of policy decisions (see, e.g., the matrix for $\firstApp$ in~\cite{FACPLTR}), the combination is not associative.

\begin{table}[!t]
\caption{The two-dimensional matrix for the $\permitOver$ combining algorithm}
\label{tab:combiningStrategies_pOver2}
$$
\begin{array}{l||c|c|c|c|}
\ \ \lfrac{d_1}{d_2}
	    & \permit & \deny & \notApp & \indet \\[.08cm]
            \hline\hline
            \permit  & \permit &  \permit &  \permit & \permit \\
            \deny & \permit  &  \deny & \deny & \indet \\
            \notApp &  \permit & \deny & \notApp & \indet \\
            \indet & \permit & \indet & \indet & \indet \\
            \hline
\end{array}
$$
\end{table}

Finally, for a generic policy $\{ \algSyntax\ \ \x{target:} \, \expr\ \ \x{policies:} \, P^{+} \, \}$, where $P^{+}$ stands for a non-empty sequence of policies or rules, its semantic clause is
$$
\!\!\policySem{\{ \algSyntax\ \ \x{target:} \, \expr\ \ \x{policies:} \, P^{+} \, \}}{\req}\,=\!\!\!\!\!\!
\begin{array}{l}
\left\{
\begin{array}{@{}l@{\ \ }l}
\effect
&  \mathtt{if} \
\policySem{\expr}{\req}=\true\ \wedge\ \policySem{\algSyntax(P^+)}{\req}= 
 \effect  \\[.1cm]
\notApp &  \mathtt{if} \
\policySem{\expr}{\req}=\false \ \vee\ \policySem{\expr}{\req}=\ \excpt \\
	& \quad\ \vee\, (\policySem{\expr}{\req}=\true \wedge \policySem{\algSyntax(P^{+})}{\req}=\notApp) \\[.1cm]
\indet &  \mathtt{otherwise}\\
\end{array}
\right.
\end{array}
$$
where $\policySem{\algSyntax(P^+)}{\req}$ is the decision returned by evaluating the combining algorithm $\algSyntax$ on the sequence of (decisions resulting from the evaluation of) policies or rules $P^+$. Thus, the policy applies to the request when the target evaluates to $\true$ and the semantic of the combining algorithm $\algSyntax$ (which is applied to the enclosed sequence of policies and the request) returns a decision $\effect$, i.e. $\permit$ or $\deny$. In this case, the resulting decision of the policy is $\effect$. Instead, if the target evaluates to $\false$ or to $\excpt$, or the combining algorithm states that the contained sequence of policies is not applicable, the policy does not apply to the request. In the remaining cases, an error has occurred and the decision is indeterminate. 

%% file: tex/security_properties.tex
% !TEX root =  ../prop_analysis.tex

\section{Security Properties for Polices}
\label{sec:sec_prop}

Policy-based specifications are sufficiently flexible and expressive to permit addressing, even in a mixed-up way, different security aspects. As stated in the Introduction, verifying whether a policy enforces a given security property is not straightforward. Therefore, in this section, we first present the attribute-based controls that the policy-based specifications must contain for ensuring various security properties. Then, we exploit the semantics of policies to formalise under which conditions a policy properly enforces such properties.

We start by providing a more precise definition of the three general security principles mentioned in the Introduction. Given a controlled system, we let $res \in Res$, $Sub' \subseteq Sub$ and $Act' \subseteq Act$, where $Res$, $Sub$ and $Act$ are respectively the set of resources, subjects and actions involved in the system's operation. Then, the three principles can be defined as follows:
\begin{itemize}
\item \emph{confidentiality}: the resource $res$ has the property of \emph{confidentiality} with respect to subjects $Sub'$ and actions $Act'$ if none of the subjects in $Sub'$ can execute actions in $Act'$ on $res$;
\item \emph{integrity}: the resource $res$ has the property of \emph{integrity} with respect to subjects $Sub'$ and actions $Act'$ if actions in $Act'$ executed by subjects in $Sub'$ cannot alter the trustworthiness of $res$;
\item \emph{availability}: the resource $res$ has the property of \emph{availability} with respect to subjects $Sub'$ and actions $Act'$ if all subjects in $Sub'$ can execute all actions in $Act'$ on $res$.
\end{itemize}

It is worth noticing that the above principles could be naively instantiated by resorting to checks on the identity of subjects. For example, when a subject whose identifier is $s$ tries to access the resource $res$, confidentiality could be achieved by denying the access to $s$ if $s \in Sub'$. However, this requires to know the identity of the requestor, as well as of all the other forbidden subjects, in advance. To overcome this limitation, different instantiations of these principles have been proposed, which rely on the features of subjects and resources for characterising the set $Sub'$ of (dis)allowed subjects. We present some of these instantiations below, by focussing on the attribute-based controls necessary for expressing the wanted features and checking specific security aspects.

Notably, from the access control point of view, the availability principle implies that the policy-based specifications have to grant the access to a subject that exhibits all the required credentials. This goal is achieved ``by construction'' in the proposed instantiations of the confidentiality and integrity principles, hence we do not further insist on the availability principle.

\subsection{Attribute-based Characterisation}
\label{sec:prop}

We use attribute names of the form $\x{subject/*}$, $\x{actions/*}$ and $\x{resource/*}$ to identify the characteristics of a \emph{subject} willing to perform a given \emph{action} on a \emph{resource}. For example, for a given access tentative, $\x{action/id}$ returns the identifier of the requested action, like e.g. \emph{read} or \emph{write}.

In the following attribute-based characterisation of the security properties, we rely on the commonly used \emph{close-world} assumption~\cite{VimercatiFS08} of access control systems, which forbids all behaviours that are not explicitly granted. We show in Section~\ref{sec:tool} how this assumption can be enforced using \facpl\ policies.

\medskip
\noindent
\textbf{Confidentiality: multi-level security}.
The security policies commonly referred to as \emph{multi-level security}~\cite{Bell76, Sandhu93} represent typical instantiations of the confidentiality principle and are also the formal basis of the MAC approach. The goal of these kinds of policies is to prevent that a resource with a certain confidentiality level be disclosed to a subject with a lower level. To this aim, each subject and resource is assigned, through a function $f_L$, a confidentiality level from a given partially ordered set $<L,\leq_L>$ of levels. In our case study, this function could be thus used to define, from the  resources' point of view, the confidentiality of loan requests and, from the subjects' point of view, the trustworthiness of clerks.

The Bell-LaPadula model~\cite{Bell76} formalises these security policies in terms of some security properties that must hold with respect to \emph{read} and \emph{write} actions\footnote{For the sake of presentation, we introduce the Bell-LaPadula model with respect to \emph{read} and \emph{write} actions; the same approach can be pursued for modelling the \submit\ and \approve\ actions of the case study.}. These properties are defined as follows:
\begin{itemize}
\item \emph{no read-up}: a subject $s$ can read a resource $res$ only if the the security level of the subject dominates the one of the object, i.e. $f_L(res) \leq_L f_L(s)$;
\item \emph{no write-down}: a subject $s$ can write a resource $res$ only if the level of the subject $s$ is dominated by the level of the object, i.e. $f_L(s) \leq_L f_L(res)$.
\end{itemize}

If we let the attributes $\x{subject/level}$ and $\x{resource/level}$ denote the confidentiality level assigned by function $f_L$ to subjects and resources, respectively, then the previous properties can be characterised in terms of policy-based specifications by the following rules:
\begin{equation}
\label{rule:conf1}
\begin{array}{l}
(\permit\ \ \mathtt{target}:  \x{equal}(\x{action/id}, read)\ \x{and}\ \x{leq}(\x{resource/level}, \x{subject/level}))\\[.2cm]
(\permit\ \ \mathtt{target}:  \x{equal}(\x{action/id}, write)\ \x{and}\ \x{leq}(\x{subject/level}, \x{resource/level}))
\end{array}
\end{equation}
where function $\x{leq}$ corresponds to the partial order relation $\leq_L$. 

The Bell-LaPadula model is usually extended to also consider DAC controls. For instance, if we use access control lists as a DAC approach, these controls could be rendered by the following rule:
\begin{equation}
\label{rule:conf2}
(\permit\ \ \mathtt{target}: \x{equal}(\x{action/id},read)\ \ \x{and}\ \ \x{in}(\x{subject/id}, \x{resource/read.ids}))
\end{equation}
where we assume that the attribute $\x{resource/read.ids}$ returns the set of all subjects allowed to execute the read action on the resource.

\medskip
\noindent
\textbf{Integrity: separation of duty}.
The integrity principle regards various system aspects in addition to accesses authorisation, like e.g. the trustworthiness of conveyance and storage means used by the system to keep resources. Since we only focus on access control, we instantiate the principle in terms of the Biba model~\cite{Biba} and the property of separation of duty.

The Biba model formalises integrity with respect to execution of read and write actions in terms of integrity levels associated with subjects and resources. Assuming that the integrity levels are defined in the same way as the confidentiality ones, the Biba model is the `dual' of the Bell-LaPadula one in that it relies on the \emph{no read-down} and \emph{no write-up} properties, which can be characterised as before.

An additional property that instantiates the integrity principle is \emph{separation of duty} (SoD), which was introduced in the Clark-Wilson model~\cite{ClarkW87} and since then has been largely adopted to define secure systems. In general, this property ensures that if two or more actions are required to perform a critical transaction, then these actions must be performed by at least two different subjects. SoD is valuable in deterring fraudulent behaviours, since no single subject has the possibility to perform complex actions, but only well-defined, elementary actions.

A basic example of SoD is to prevent that an action be executed when a subject is assigned two roles that are conflicting, i.e. there is no separation of duties among the actions that these roles permit. For instance, if we assume roles $role1$ and $role2$ to be in conflict, we can define a rule that permits a write action only when a subject exposes the first role but not the second one; the rule is as follows
\begin{equation}
\label{rule:int2}
\begin{array}{l}
 (\permit\ \ \mathtt{target}:  \x{equal}(\x{action/id}, {write})\ \x{and}\\
 \qquad\qquad\qquad\quad\ \ \x{in}(role1, \x{subject/role})\ \x{and}\ \x{not}(\x{in}(role2,\x{subject/role})))
 \end{array}
\end{equation}
Indeed, the rule checks that the roles assigned to the subject, that are obtained through the attribute $\x{subject/role}$, include $role1$, which is required for executing the read action, and not $role2$. In the case study, by using the rule above where $\emph{write}$, $\emph{role1}$ and $\emph{role2}$ are replaced by, respectively, \approve, \roleOne\ and \roleTwo, we can ensure that a clerk assigned with both roles cannot perform an \approve\ action.

This last property is an example of \emph{static} SoD, i.e. an integrity requirement that can be fulfilled by evaluating a single access request. However, if we define SoD in terms of conflicting actions, rather than in terms of conflicting roles as done before, checking a single access request is not adequate anymore to enforce the intended property. Indeed, SoD could be easily circumvented by executing conflicting actions in two or more attempts, as it is the case of a subject that is assigned, in two different instants of time, different roles granting conflicting actions. To avoid that the subject be authorised to execute both actions, we must resort to considering the previous actions it has performed, which is an example of \emph{dynamic} SoD. This kind of properties can be still addressed by using policy-based specifications, but we need to use attributes for storing the history of the accesses previously performed by a subject. We will provide further details on this aspect when discussing future work.

\medskip
\noindent
\textbf{Role-based design: hybrid properties and least-privilege}. The role-based design is a high level approach that permits to enforce confidentiality and integrity properties on the controlled resources at the same time. It consists in assigning different roles to subjects within the system and using policies stating what accesses are allowed to subjects depending on the roles they have. Although it is not an instantiation of one of the three general security principles, we consider this approach explicitly since it is largely used due to its better scalability with respect to other models, like e.g. the Bell-LaPadula and Biba ones.

The basic characterisation of role-based controls in terms of policy-based specifications is straightforward: the attribute $\x{subject/role}$ permits to define controls on the subject's roles and Rule~(\ref{rule:int2}) is a concrete example of this. However, the role-based approach takes also different, more complicated forms~\cite{SandhuCFY96}, that exploit role hierarchies, i.e. a role inherits the privileges of the roles that are higher up in the hierarchy, or constraints on role assignments, i.e. two conflicting roles cannot be assigned at the same time. The former case can be rendered by exploiting the hierarchy of polices or an appropriate ordering function, while the latter one by using an approach similar to that used for SoD properties. The characterisation of role-based controls thus formalises a sort of `hybrid' property, consisting of both confidentiality and integrity aspects.

Let us consider an hybrid property stating that an action $write$ can be executed by all the subjects with assigned role $role3$, or by any other subject in the underlying role hierarchy which is not assigned role $role4$ at the same time. Its characterisation in terms of attribute-based specifications can be defined as follows:
\begin{equation}
\label{rule:role1}
\begin{array}{l}
 (\permit\ \ \mathtt{target}:  \x{equal}(\x{action/id}, write)\ \x{and}\\
\qquad\qquad\qquad\quad\ \
 \x{sub}\textrm{-}\x{role}(\x{subject/role}, role3)\
\x{and}\ \x{not}(\x{in}(role4,\x{subject/role})) )
\end{array}
\end{equation}
where we use the ad-hoc function $\x{sub}\textrm{-}\x{role}$ to check if the subject's role is a sub-role of (or coincides with) $role3$ and the additional control on role $role4$ to encode the integrity check.

A guideline commonly used in role-based design is \emph{least privilege}. It means that each subject should not expose more privileges than those necessary to perform the requested action. Differently from the properties we have previously considered, least privilege is not implemented through specific rules. Rather, it affects the design choices pursued for defining access control policies. We will present more details in the semantic-based formalisation presented in the next section.

\subsection{Semantic-Based Formalisation}
\label{sec:formal_prop}

A policy-based specification, as e.g.~a \facpl\ policy, in addition to the rules previously presented, contains many other elements, such as e.g. other rules implementing additional controls and conflict resolution strategies. We now formalise under which conditions a policy enforces a given security property.

The formal representation of a security property is obtained by exploiting the fact that an access control request is an assignment of values to a collection of attributes. We can then use sets of requests to represent the (non)secure system behaviours with respect to a given property. Formally, given a security property $\mathit{pr}$, we let $\Req_{pr}$ (resp., $\overline{\Req}_{pr}$) be the \emph{permit} (resp. \emph{deny}) set, i.e. the set of requests that represent the secure (resp., nonsecure) behaviours with respect to $\mathit{pr}$, and $Sub_{pr}$ (resp., $Res_{pr}$) be the subset of subjects (resp., resources) for which the property $pr$ is defined. A policy $P$ containing the rules characterising $\mathit{pr}$ correctly enforces such property if the following conditions hold
$$
\begin{array}{l@{\  }l@{\ }l@{\ }l@{\ }l}
\forall\ \req\ \in \Req_{pr} & : \ \req\,(resource/id) \in Res_{pr}\,,\, \req\,(subject/id) \in Sub_{pr} &  \Rightarrow \ \policySem{P}{\req} &=\permit & \\[.1cm]
\forall\ \req\ \in \overline{\Req}_{pr} & : \ \req(resource/id) \in Res_{pr}\,,\, \req(subject/id) \in Sub_{pr} & \Rightarrow \  \policySem{P}{\req}& =\deny &
\end{array}
$$
where notation $\req(\x{attr\_name})$ indicates the value assigned to the attribute named $\x{attr\_name}$ by the request $\req$. Hence, we require that all the secure behaviours are allowed (i.e., all the requests in $\Req_{pr}$ evaluate to $\permit$) and all the nonsecure ones are forbidden (i.e., all the requests in $\overline{\Req}_{pr}$ evaluate to $\deny$). Notably, we consider the (non)secure behaviours that only refer to the subset of subjects and resources that the property $pr$ takes into account. This means that the set $\overline{\Req}_{pr}$ is not the complementary set of $\Req_{pr}$ with respect to the universe of all the possible behaviours of the system; rather it represents those behaviours that are considered nonsecure by the property $pr$. In the sequel we report the definition of the (non)secure sets of requests for each property we presented before.
%
%\todo{Review 2: "I would formalise the security properties as properties of the semantic domain ..." Questa parte non mi pare si possa considerare. Credo si riferisca pi\`u a cose tipo ``secure" information-flow garantito da un type-system e simili, qui non vedo come mappare le propriet\`a considerate sul semantic domain\\[.2cm] FT: ok, lasciamo come \`e (magari pi\`u avanti proviamo a ripensarci per vedere se riusciamo a capire meglio...)\\[.2cm] Ros: d'accordo, pensiamoci con calma, magari per i post proceedings. }

\medskip
\noindent
\textbf{Confidentiality: multi-level security}. The secure behaviours identified by the \emph{no read-up} property corresponds to the set of requests $\Req_{nru}$ whose elements $\req$ must satisfy the following conditions
$$
\req
(\x{action/id})= read\ ,\, \req
(\x{resource/level})= l_1 \,,\, \req
(\x{subject/level})=l_2 \quad : \quad  l_1, l_2 \in L\ ,\ \  l_1 \leq_L l_2
$$
The set $\overline{\Req}_{nru}$ instead contains those requests satisfying the following conditions
$$
\req
(\x{action/id})= read\ ,\, \req
(\x{resource/level})= l'_1 \ , \, \req
(\x{subject/level})=l'_2 \quad : \quad  l'_1, l'_2 \in L\ ,\ \  l'_1 \not\leq_L l'_2
$$
The permit and deny sets for the \emph{no write-down} property are similarly defined. In case of DAC properties as e.g. that defined by Rule~(\ref{rule:conf2}), the requests of the set $\Req_{dac}$ are characterised by the following conditions
$$
\req
(\x{action/id})= read\,,\, \req
(\x{resource/read.ids})= Sub_{res} \,,\, \req
(\x{subject/id})= s \quad : \quad  s \in Sub_{res}
$$
where $Sub_{res}$ is set of all subjects allowed to execute the read action on the resource ${res}$. Instead, the elements of the deny set $\overline{\Req}_{dac}$ must satisfy the following conditions
$$
\req
(\x{action/id})= read\,,\, \req
(\x{resource/read.ids})= Sub'_{res} \,,\, \req
(\x{subject/id})=s \quad : \quad
s \not\in Sub_{res}'
$$
Indeed, the set of granted subjects $Sub'_{res}$ does not contain the subject $s$.

\medskip
\noindent
\textbf{Integrity: separation of duty}. The \emph{no read-down} and the \emph{no write-up} properties, representing the Biba model, are formalised like the confidentiality ones. 

Let us consider the SoD property for a {write} %read 
action expressed by Rule~(\ref{rule:int2}). Thus, if $\emph{Rol}$ is the set of authorised non conflicting sets of roles, i.e. all the sets contain $role1$ and not $role2$, the secure behaviours $\Req_{sod}$ are defined as follows
$$
\req
(\x{action/id})= %read
{write} \ ,\ \req
(\x{subject/role})= rol \quad : \quad rol \in \mathit{Rol}
$$
The non secure behaviours $\Req_{sod}'$ are instead defined as follows
$$
\req
(\x{action/id})= %read
{write} \ ,\ \req
(\x{subject/role})= rol' \quad : \quad rol' \in Rol_{all} \backslash \mathit{Rol}
$$
where the set $Rol_{all}$ represents the set of all sets of roles that a subject can play in the system. Thus, a request is non secure when the set of subject's roles does not contain $role1$, i.e. the subject has not the right to execute the %$read$
{$write$} action, or it contains $role1$ and $role2$ at the same time, i.e. the exposed roles are in conflict.

\medskip
\noindent
\textbf{Role-based design: hybrid properties and least-privilege}.
The secure and nonsecure behaviours identified by hybrid properties are just a combination of the previous examples. The formalisation of the least privilege requires instead additional comments.

Let us consider a security property $pr$ and the set of request $\Req_{pr}$ representing the secure behaviours with respect to such property. The sets of secure and nonsecure behaviours for the least privilege, with respect to $pr$, are defined as follows
$$
\Req_{lp} = \Req_{pr}\qquad\qquad \overline{\Req}_{lp} = \Req_{all} \backslash \Req_{pr}
$$
where $\Req_{all}$ indicates all the possible requests. Therefore, in order to enforce the least privilege, a policy has to authorise all those behaviours of the system deemed as secure by the property $pr$ and to forbid all the other behaviours, not only those violating $pr$ as in the previous cases. All the behaviours that are not defined secure by $pr$ are considered as nonsecure. Hence, forbidding them ensures that possibly granted accesses cannot be used to circumvent, in a malicious way, other policies in the system.

%% file: tex/struct_prop.tex
% !TEX root =  ../prop_analysis.tex

\section{Structural Properties for Policies}
\label{sec:struct}

We now address some of the properties proposed in the literature which refer to the structure of policies. We start considering completeness of a single policy, after which we will consider redundancy, disjointness and coverage of one policy with respect to other ones. The properties dealing with multiple policies capture the relationships among the different sets of system behaviours they enforce. In this section, we report a uniform characterisation of these properties by means of the semantic-based approach used before.

By referring to \facpl, we use $P$ to range over policies, $\mathsf{alg}$ to range over combining algorithms and $d$ to range over authorisation decisions. Moreover, we use $\Req_{all}$ to denote the set of all possible requests.

\medskip
\noindent
\textbf{Completeness}. A policy $P$ is \emph{complete} if there is no access request for which there is an absence of decision. Formally, this property can be rendered through the following condition
$$
\forall\ \req\ \in \Req_{all} : \ \policySem{P}{\req} \neq \notApp
$$
In fact, we require that the policy applies to any request, i.e. it always returns a decision different from $\notApp$. Notably, in this formulation $\indet$ is considered as an admissible decision; a more restrictive formulation could be defined that only accepts decisions $\permit$ and $\deny$.

\medskip
\noindent
\textbf{Redundancy}. Redundancy among policies means that to enforce the same set of system behaviours some policies are not needed. Therefore, if we eliminate redundant policies, we can improve performance of policy evaluation while leaving unchanged the enforced behaviours. Although the concept seems natural and quite simple, different formalisations, that often lack of precision, have been proposed in the literature. We follow an approach similar to~\cite{GuarnieriNMM13}.

Formally, if we let the \facpl\ policy $S$ be defined as $S = \mathsf{alg}(P_1, \ldots,P_i, P, P_{i+1}, \ldots, P_n)$, then the policy $P$ is \emph{redundant} with respect to $S$ if the following condition holds
$$
\begin{array}{l}
\forall\ r \in \Req_{all} : \ \policySem{\mathsf{alg}(P_1, \ldots,P_i, P, P_{i+1}, \ldots, P_n)}{\req} \ = \  \policySem{\mathsf{alg}(P_1, \ldots,P_i, P_{i+1}, \ldots, P_n)}{\req}
\end{array}
$$
In fact, we require that, for any request, the decision returned by $S$ is not affected by the presence of $P$. Notably, this property generalises in the obvious way to the case $S$ contains rules instead of policies (thus $P$ would be a redundant rule) and to the case a target is present in $S$.

\medskip
\noindent
\textbf{Disjointness}. Disjointness among policies means that such policies apply to disjoint sets of behaviours. Thus, two policies are \emph{disjoint} if there is no request for which both policies evaluate to $\permit$ or $\deny$. Formally, policies $P$ and $P'$ are \emph{disjoint} if the following condition holds
$$
\begin{array}{l}
\forall\ r \in \Req_{all} \ \ : \ \ \{ \: \policySem{P}{\req}, \ \policySem{P'}{\req} \: \} \not\subseteq \{\permit, \deny\}
\end{array}
$$
It is worth noticing that disjoint polices can be combined with the assurance that the allowed or forbidden behaviours enforced by each of them are not in conflict, which simplifies the choice of the combining algorithm to be used.

\medskip
\noindent
\textbf{Coverage}.
Coverage among policies means that one of such policies enforce the same decisions as the other ones for a set of requests of interest. Formally, if $\Req_{cov}$ is a set of requests, we say that the policy $P$ \emph{covers} the policy $P'$ if, for each request $\req \in \Req_{cov}$ to which $P'$ applies, i.e. $\policySem{P'}_\req \in \{\permit, \deny\}$, $P$ applies too and returns the same decision. Formally, it is expressed by the following condition
$$
\begin{array}{l}
\forall\ r \in \Req_{cov} \ \ : \ \  \policySem{P'}{\req} \in \{\permit, \deny\} \ \ \Rightarrow\ \ \policySem{P}{\req} = \policySem{P'}{\req}
\end{array}
$$
Thus, relatively to the set of requests of interest, $P$ enforces at least the same allowed and forbidden behaviours as $P'$. Consequently, if $P'$ also covers $P$, then the two policies enforce exactly the same behaviours relatively to the set of requests of interest.

\medskip

These structural properties permit to statically reason on the relationships among policies and provide useful support to system's designers in developing and maintaining policy-based specifications. One technique they support is the \emph{change-impact analysis}~\cite{FislerKMT05}. This analysis examines the effect of policy modifications for discovering unintended consequences of such changes. To be practically effective it requires that the verification of the previous properties be supported by automatic tools. We further deal with this issue in the next section.

%% file: tex/constraint.tex
% !TEX root =  ../prop_analysis.tex

\section{Verification of Properties}
\label{sec:tool}

The formalisation of security and structural properties presented in Sections~\ref{sec:sec_prop} and~\ref{sec:struct} determines the conditions on attributes stating when a policy enjoys a certain property. To verify such conditions, we need to take into account the various elements composing a policy. Specifically, the hierarchical structure of policies and the various elements originating the decisions make this verification cumbersome and error-prone if not supported by an automatised technique. As an example of the difficulties to be faced, we consider (part of) the policies modelling the case study, which address the \emph{no read-up} and DAC security properties for read actions requested by a set of subjects $Sub'$ on the resource $\doc$. Thus, we define various combination approaches for creating a policy containing Rules~(\ref{rule:conf1}) and~(\ref{rule:conf2}), and we study for each approach if the two properties are properly enforced.

The first combination we propose for the two rules is defined as follows
$$
\begin{array}{l}
\{ \permitOver \\
\ \ \mathtt{target}: \x{equal}(\x{resource/id}, {\doc})\ \x{and}\ \x{in}(\x{subject/id}, Sub')\\
\ \ \mathtt{policies}: \\
\qquad (\permit\ \ \mathtt{target}:  \x{equal}(\x{action/id}, read)\ \x{and}\ \x{leq}(\x{resource/level}, \x{subject/level}))\\
\qquad (\permit\ \ \mathtt{target}: \x{equal}(\x{action/id},read)\ \x{and}\ \x{in}(\x{subject/id}, \x{resource/read.ids}))
\}
\end{array}
$$
The chosen combination algorithm is $\permitOver$, which seems the natural choice since each allowed behaviour is explicitly authorised. Notably, the policy's target ensures that the policy exclusively applies to the considered resource {$\doc$} and to the subset of system's subjects $Sub'$.

To verify that this policy enforces the intended properties, we show that all the secure behaviours are authorised, while the nonsecure ones are forbidden. We consider first the \emph{no read-up} property. As formalised in Section~\ref{sec:formal_prop}, the secure behaviours correspond to all the requests containing the resource and subject levels that respect the partial ordering relation. These ones clearly match the target of the first rule, hence this rule,  as well as the $\permitOver$ algorithm, return $\permit$. The nonsecure behaviours are instead represented by all the requests containing resource's and subject's levels not properly ordered. In this case, both internal rules do not apply and the $\permitOver$ algorithm returns $\notApp$, because neither $\permit$ nor $\deny$ are returned by the rules. However, the nonsecure behaviours should be evaluated as $\deny$, hence we can conclude that the policy does not properly enforce the \emph{no read-up} property. The same also holds for the DAC property.

To fix this first policy, we can replace the $\permitOver$ algorithm by the $\denyUnless$ one, which ensures that $\deny$ is taken as the default decision whenever no rule evaluates to $\permit$. In this case all the nonsecure behaviours of both properties are properly forbidden. However, as we are addressing two properties, the secure behaviours are all those ones that are secure, at the same time, for both properties. This means that $\permit$ must be returned only when the two rules apply at the same time as well, but this does not happen in the presented policies. In fact, the combining algorithm does not enforce any form of consensus between the two rules. As a matter of fact, a subject can circumvent the access control system reading a resource, e.g., only having the correct confidentiality level and not the discretionary access.

This additional issue can be addressed by adding a new policy layer and requesting a strong consensus between the rules. The extended policy is thus as follows
$$
\begin{array}{l}
\{
\denyUnless\\
\ \ \mathtt{policies}:\\
\ \ \ \{ \strongCon \\
\ \ \ \quad \mathtt{target}: \x{equal}(\x{resource/id}, {\doc})\ \x{and}\ \x{in}(\x{subject/id}, Sub')\\
\ \ \ \quad \mathtt{policies}: \\
\ \ \ \qquad (\permit\ \ \mathtt{target}:  \x{equal}(\x{action/id}, read)\ \x{and}\ \x{leq}(\x{resource/level}, \x{subject/level}))\\
\ \ \ \qquad (\permit\ \ \mathtt{target}: \x{equal}(\x{action/id},read)\ \x{and}\ \x{in}(\x{subject/id}, \x{resource/read.ids}))
\  \ \ \} 
\}
\end{array}
$$
$\denyUnless$ is used at top level to ensure that the resulting decisions of the overall policy will be only $\permit$ or $\deny$. In the inner policy, $\strongCon$ ensures that $\permit$ is returned only when both internal rules apply at the same time. In this case, all secure and nonsecure behaviours of the two intended properties are properly enforced. Notably, we can achieve the same result by merging the two rules and avoiding the additional policy layer; however, the modelling approach we present permits to achieve separation of concerns among rules, which are thus easier to maintain and possibly change.

Verifying that a policy properly enforces a set of properties is not straightforward. This example, which seems easy enough for being manually checked, shows us that also in case of simple policies we need an automated verification approach. Specifically, this approach must be capable to take into account all the aspects of a policy specification, e.g. policy stratification and combining algorithms, and to exhaustively check all the significant requests representing the possible behaviours. A viable approach towards an automated verification of security and structural properties is outlined in the next subsection.

\subsection{Towards an Automated Verification Approach}
\label{sec:aut_ver}

Automatising the verification of properties permits to facilitate the analysis of policy-based specifications. To enable such analysis, we need a formalism that, on the one hand, permits to collapse hierarchical policies into a single-layered representation and to uniformly represent all policy elements and, on the other hand, is sufficiently flexible to deal with multiple domain values for attribute assignments. To this aim, we propose a constraint-based formalism.

Constraints permit to specify satisfaction problems based both on boolean formulae and on formulae dealing with different theories as, e.g. linear arithmetics. Such kind of formulae are called \emph{satisfiability modulo theories} (SMT) formulae. Choosing an SMT-based approach is advocated also by the relevant progress made in the development of automatic SMT solvers (e.g., Z3~\cite{MouraB08}), which make SMT formulae to be extensively employed in diverse analysis applications~\cite{MouraB11}. In addition, the analysis of logic-based access control policies reported in~\cite{ArkoudasCC14} points out that the SMT-based approach is more effective than many other ones, like e.g. the approaches based on decision diagrams~\cite{FislerKMT05} or on description logic~\cite{KolovskiHP07}. Of course, the feasibility of the SMT-based approach crucially depends on decidability of the satisfiability checks; in other words, the used constrains must be represented by decidable theories, as e.g. uninterpreted functions and array theories.

To achieve a single-layered representation of policies, we have to provide a translation function from the language used for writing policies to the constraint-based formalism that preserves the semantics of the original language. Indeed, since \facpl\ is equipped with a formal semantics, it has to be exploited for defining a rigorous encoding. Notably, as the evaluation of a policy can return four possible decisions, we have to define a different constraint for each of them.

A constraint-based representation of policy-based specifications enables the verifications of both security and structural properties. Specifically, in the case of security properties, the attribute values identifying the class of (non)secure requests correspond to assignment assertions in the constraint of interest (i.e. the one modelling the decision to which the requests should evaluate) and then, by means of an SMT-solver, it is checked if such constraint is satisfiable. If this happens, it means that the requests of the class can evaluate, under the assignment model returned by the solver, to the decision modelled by the constraint. In case of structural properties, we can instead define boolean combinations among the single constraints of each policy, and then check the satisfiability of the resulting constraint to understand if a certain property holds. For instance, the disjointness between two policies holds if the constraint resulting from the implication of the $\permit$ (resp., $\deny$) constraints of both policies is not satisfiable.

%% file: tex/relatedworks.tex
% !TEX root =  ../prop_analysis.tex

\section{Related Works}
\label{sec:relwork}

%%%%%%%%%%%%%%%%%%%%%%%%%
%Languages for policies
%%%%%%%%%%%%%%%%%%%%%%%%%
Policy-based specifications have recently been the subject of extensive research, both by industry and academia, in many application areas. In fact, policy languages have been adopted for managing different aspects of systems' behaviour, not only access control but also adaptation enforcement and network management. A large variety of languages for defining access controls has been proposed, and the more significant ones follow two main specification approaches: \emph{rule-based}, as e.g. XACML~\cite{XACML3} and Ponder~\cite{DamianouDLS01}, and \emph{logic-based}, as e.g. ASL~\cite{Jajodia97alogical} and the logical framework presented in~\cite{ArkoudasCC14}. We present the relevant features of these languages, showing the effectiveness of choosing \facpl\ as the target language for studying policies' properties. Notably, the uniform approach based on attributes presented in~\cite{JinKS12} does not provide any evaluable property characterisation, but only an high-level access control model.
%
%\todo{\scriptsize AM: Richiedono un confronto pi\'u dettagliato e profondo per la parte di analisi quando si dice che SMT \`e migliore\\[.2cm] FT: ok, per lasciamo cos\`i e poi vediamo per i post-proceedings\\[.2cm] Ros: d'accordo, rimandiamo ai post proceedings\\  AM: aggiunta frase in sezione 5.1}\

XACML is the most widely-used instantiation of the ABAC approach. It relies on an XML-based syntax and permits to write policies and access requests. However, XML does not permit compact specifications and, due to the lack of a formal semantics, an explicit unambiguous formalisation of request's evaluation. The use of \facpl\ permits thus to avoid verbose examples, and to rely on a rigours formal semantics to formalise properties.

Ponder is instead a strongly-typed language defined in terms of Event-Condition-Action rules. Differently from XACML and \facpl, it does not provide any explicit combination strategy to resolve conflicts. Thus, the presence of conflicts or inconsistency is statically analysed by means of abductive reasoning techniques~\cite{BandaraLR03}. This reasoning generates a refinement for the considered policy. Ponder, on the one hand, permits to avoid policy hierarchies, but, on the other hand, it does not provide any modularity and compositionality in the specification of policies. The \facpl -based specification approach consists instead in basic building rules, that can be appropriately combined to enforce different security properties, ensuring separation of concerns in the enforced behaviours.

%%%%%%%%%%%%%%%%%%%%%%%%%
%Property Verification and Encoding
%%%%%%%%%%%%%%%%%%%%%%%%%
The increasing spread of policy-based specifications has prompted the development of multiple verification techniques like, e.g., property checking and behavioural characterisations. Such techniques have been implemented by means of different formalisms, varying from multi-terminal binary decision diagrams (MTBDD) to different kinds of logics.  We review the more relevant techniques and formalisms.

The change-impact analysis of XACML policies presented in~\cite{FislerKMT05} permits to study the consequences of policy's modifications. In particular, to verify structural properties among policies by means of automatic tools, this approach relies on a MTBDD-based representation of policies. However, it cannot deal with many of the classical combining algorithms, e.g. all the XACML's ones,  and, as outlined in~\cite{ArkoudasCC14}, an SMT-based approach (i.e. the one we are exploring), scales significantly better than the MTBDD one.

The ASL language~\cite{Jajodia97alogical} is a logical framework for the formalisation of access control policies. Specifically, it enables hierarchisation, conflict resolution, and role- and group-based definitions of access rights. Furthermore, by means of additional predicates representing a posteriori checks on authorisation decisions, it permits to easily express various history-dependent properties, \eg dynamic separation of duty.
Similarly, the framework in~\cite{ArkoudasCC14} permits a logic-based specification of control policies. A policy is thus a list of constraint assertions that are evaluated by a SMT-based tool, and various structural properties can be encoded in terms of additional, low-level assertions.
The \facpl -based approach permits instead to abstract from the underlying logical means (that are still used to in the \facpl\ formal semantics and for the automatised analysis we foster), allowing a better usability for system's designers of the properties formalisation.

An additional logic-based analysis is the one presented in~\cite{KolovskiHP07}, which aims at verifying structural properties of XACML policies. Specifically, it defines a partial encoding of XACML into description logics and a set of supporting analysis services. However, this approach does not take into account many combining algorithms and, also, the decisions $\notApp$ and $\indet$, which are instead useful in the definition of structural properties. Furthermore, the used reasoning tool suffers the same scalability issues as the one based on MTBDD.

Finally, the redundancy property has been object of specific intensive studies. In fact, the identification of redundant policies and their `safe' elimination increases the evaluation performance of access control systems. A rigorous formalisation of redundancy is proposed in~\cite{GuarnieriNMM13}, where an algorithmic approach for minimising access control policies is proposed and its computational complexity studied.

%% file: tex/conclusion.tex
% !TEX root =  ../prop_analysis.tex

\section{Conclusion}
\label{sec:concl}

Policy-based specifications are widely used to regulate the behaviour of system's entities relatively to the access to shared resources. The policy-based access control, by resorting to the concept of attribute, is sufficiently expressive to represent in an uniform way all classical access control approaches, varying from access control list and role-based to discretionary and mandatory ones. Policies permit indeed to define fine-grained, flexible and context-aware access controls, fostering systems integration, as attributes can be retrieved from different information systems. To ensure confidentiality and integrity principles, such policies need to take into account multiple security aspects, e.g., the ones studied by well-known security models, such as the Bell-LaPadula and Biba ones. However, enforcing in terms of policy-based specifications the security properties characterising such models is a tricky task. In fact, the hierarchical structure of policies, the presence of conflict resolution strategies and the intricacies deriving from the many controls involved do not permit to easily check whether a given security property is properly enforced. By means of the \facpl\ policy language, we have provided some specification examples of a significant set of security properties, and showed under which conditions such properties are properly enforced. To characterise the relationships with the behaviours that different polices enforce, we have also formalised, in a uniform way, various properties on the structure of policies. Furthermore, to effectively support system's designers in developing and maintaining policy-based specifications, we outlined a constraint-based approach enabling automated verification of security and structural properties by means of constraint solver tools.

\label{sec:futureworks}

We conclude by reviewing some additional properties we plan to study in the next future. On the one hand, to take into account dynamic behaviours of systems, we want to address history-dependent security properties, and provide specialised formal analysis techniques. On the other hand, access control policies can also be used to produce, together with the authorisation decision, additional actions, named \emph{obligations}, that can adapt the computing system's configuration. To reason on obligations, we want to formalise properties on conflicts and dependencies among them. Further details follow.

\medskip
\noindent
\textbf{History-Dependent Properties}.
Classical examples of history-dependent properties are dynamic SoD and Chinese Wall~\cite{BrewerN89}. Dynamic SoD properties correspond to enforcing separation of duty by evaluating not only the current subject's request, but also the history of actions the subject has previously performed. Chinese Wall properties correspond instead to an hybrid instantiation of the confidentiality and integrity principles, where history is used to adapt the access rights granted by the confidentiality controls. Specifically, it means that a subject is only allowed to access resources which are not in conflict with any other resource that the subject has already accessed.

Enforcing these properties within policy-based specifications means checking the history of system's authorisations. This could be done, e.g., by means of attributes representing the history. These attributes should in fact collect all the information needed for properly enforcing a considered history-dependent property, e.g., in case of Chinese Wall, which resources have been already accessed. In order to formally verify that such properties are enforced, we need to enhance our semantic-based formalisation with an explicit representation of history. Possible approaches to pursue for achieving this formalisation are those used in Usage Control~\cite{LazouskiMM10}, i.e. a novel access control model for ensuring continuous authorisation when an access is in progress.

\medskip
\noindent
\textbf{Obligations}. Obligations have been introduced in access control for modelling the need of fulfilling additional actions in order to gain access. For instance, XACML supports the definition of obligations and, to allow an access, it requires that all obligations possibly generated by the policy evaluation are correctly fulfilled. Obligations can be thus used to adapt the computing system's configuration. However, these obligations may have conditional requirements on their execution, e.g. conflicts and dependencies, that have to be taken into account. For instance, an obligation can require to be executed only if another one has not been already executed. To formalise and analyse properties on obligations, we plan to start from the representation model of obligation's features outlined in~\cite{BertinoBCCKK09}, and instantiate such model with respect to the \facpl\ policy language.